%% file: main.tex
\documentclass[9pt,conference]{IEEEtran}
\usepackage{dcase2025}

\usepackage{bm} 

\usepackage{acronym}
\acrodef{ae}[AE]{autoencoder}
\acrodef{auc}[AUC]{area under the \ac{roc} curve}
\acrodef{asd}[ASD]{anomalous sound detection}
\acrodef{knn}[kNN]{$k$-nearest neighbor}
\acrodef{lora}[LoRA]{low-rank adaptation}
\acrodef{mlp}[MLP]{multilayer perceptron}
\acrodef{pauc}[pAUC]{partial \ac{auc}}
\acrodef{roc}[ROC]{receiver operating characteristic}
\acrodef{ssl}[SSL]{self-supervised learning}
\acrodef{sota}[SOTA]{state-of-the-art}

\usepackage{dcase2025,amsmath,graphicx,url,times,booktabs, tabularx}
\input{settings.tex}

\title{ASDKit: A Toolkit for Comprehensive Evaluation of \\Anomalous Sound Detection Methods}

\name{
   Takuya Fujimura$^{1}$, 
   Kevin Wilkinghoff$~^{2,3}$,
   Keisuke Imoto$^{4}$,
   Tomoki Toda$^{1}$
}
\address{
$^{1}$Nagoya University, Japan, $^{2}$Aalborg University, Denmark, $^{3}$Pioneer Centre for AI, Denmark, $^{4}$Kyoto University, Japan
}

\begin{document}

\maketitle

\begin{abstract}
In this paper, we introduce ASDKit, a toolkit for \ac{asd} task.
Our aim is to facilitate ASD research by providing an open-source framework that collects and carefully evaluates various \ac{asd} methods.
First, ASDKit provides training and evaluation scripts for a wide range of \ac{asd} methods, all handled within a unified framework.
For instance, it includes the autoencoder-based official DCASE baseline, representative discriminative methods, and self-supervised learning-based methods.
Second, it supports comprehensive evaluation on the DCASE 2020--2024 datasets, enabling careful assessment of \ac{asd} performance, which is highly sensitive to factors such as datasets and random seeds.
In our experiments, we re-evaluate various \ac{asd} methods using ASDKit and identify consistently effective techniques across multiple datasets and trials.
We also demonstrate that ASDKit reproduces the state-of-the-art-level performance on the considered datasets.
\end{abstract}

\acresetall  


\section{Introduction}
\label{sec:intro}
\Ac{asd} is a technique for machine condition monitoring, where \ac{asd} systems aim to detect mechanical failures from their operational machine sounds~\cite{Koizumi_DCASE2020_01,kawaguchi2021description,dohi2022description,dohi2023description,nishida2024description}.
In this task, since it is infeasible to exhaustively collect rare and diverse anomalous sounds, the system is developed with only normal machine sounds.
Therefore, \ac{asd} systems calculate anomaly scores based on the deviation from the normal sound distribution and assign anomalies to sounds with high anomaly scores.

The \ac{asd} field has advanced through the development of various methods.
One major approach is based on generative modeling~\cite{Koizumi_DCASE2020_01,suefusa2020anomalous,dohi2021flow}, where it directly models the distribution of audio features belonging to normal machine sounds and computes anomaly scores based on the negative likelihood of a given observation.
For example, an \ac{ae}-based method~\cite{Koizumi_DCASE2020_01} trains an \ac{ae} network to reconstruct audio features and computes anomaly scores based on the reconstruction error for a given observation.
Other variants of \acp{ae} only reconstruct the center frame of consecutive spectrogram frames \cite{suefusa2020anomalous} or mask the input to obtain an autoregressive model \cite{giri2020group}.
Another recent approach first projects audio signals into a lower-dimensional feature space and then computing anomaly scores based on the distance of an observation from the normal training data samples in that space~\cite{Wilkinghoff2023design,wilkinghoff2024self,saengthong2025deep,zheng2024improving}.
For example, \ac{sota} discriminative methods~\cite{lopez2020speaker,giri2020self,primus2020anomalous,Wilkinghoff2023design,zheng2024improving,KuroyanagiSerialOE} train a discriminative feature extractor to classify meta-information labels associated with normal training data, and then compute anomaly scores in the discriminative feature space.
Possible choices of meta information include the machine type, machine IDs and specific operating conditions of machines.
Alternatively, \ac{ssl} feature spaces of models trained on (large) external datasets are directly utilized for \ac{asd}~\cite{Wilkinghoff2023pretrained,saengthong2025deep}.

Although the proposal of various \ac{asd} methods has undoubtedly advanced the field, we argue that further development is hindered by the lack of comprehensive and careful evaluation.
A representative effort for evaluating \ac{asd} methods is the DCASE challenge~\cite{Koizumi_DCASE2020_01,kawaguchi2021description,dohi2022description,dohi2023description,nishida2024description}, which plays an important role in enabling fair evaluations using unrevealed test data.
However, challenge evaluations are conducted on a single dataset and a single trial, which is insufficient for thorough assessment, as we have observed that \ac{asd} performance is highly sensitive to datasets and random seeds.
Furthermore, it is difficult to precisely identify the effectiveness of individual components, as participants develop methods on their own frameworks and often employ ensembles of multiple systems to boost performance.

\begin{figure}[t]
   \centering
   \centerline{\includegraphics[width=0.9\columnwidth]{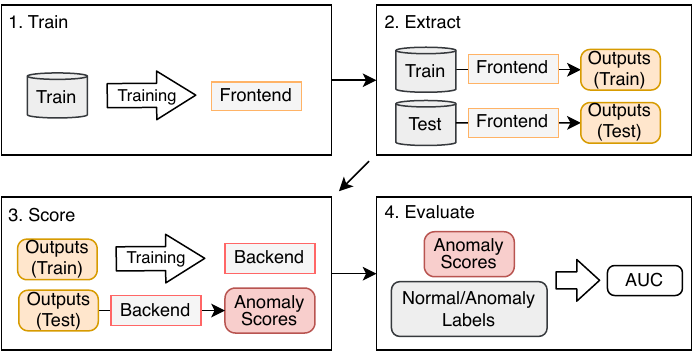}}
\vspace{-4pt}
   \caption{Unified framework of ASDKit.
(1) Training the frontend with the training dataset,
(2) extracting and storing outputs from both the training and test datasets using the frontend,
(3) computing anomaly scores for the test data using the backend trained on the training data, and
(4) computing evaluation scores based on the anomaly scores and the ground-truth normal and anomalous labels.
}
   \label{fig:framework}
\vspace{-4pt}
\end{figure}

To address these limitations, we introduce ASDKit\footnote{https://github.com/TakuyaFujimura/dcase-asd-toolkit}, an open-source toolkit for the \ac{asd} task.
The main features of ASDKit are summarized as follows:
(1) ASDKit collects various \ac{asd} methods into a single open-source repository, enabling easy access and comparison.
(2) ASDKit provides \textit{recipes} that complete the entire training and evaluation process shown in Fig.~\ref{fig:framework}.
(3) ASDKit supports comprehensive evaluation on the DCASE 2020--2024 datasets~\cite{Koizumi_DCASE2020_01,kawaguchi2021description,dohi2022description,dohi2023description,nishida2024description} with multiple trials, enabling careful assessment.
To demonstrate ASDKit's capabilities, we re-evaluate various \ac{asd} methods across multiple datasets and trials.
By analyzing the results, we identify consistently effective techniques and provide new insights.
Furthermore, we demonstrate that ASDKit can reproduce the \ac{sota}-level performance on the considered datasets.

\section{ASDKit}

\begin{table*}[t]
   \centering
   \caption{Supported experimental conditions in ASDKit.
   MT and Sec denote machine type and section, respectively.
   The ``\#ID/Sec'' column shows the number of machine IDs or sections within each machine type.
   The ``Dev/Eval'' column shows the information used to define the dev and eval subsets. 
   The ``Aggregation'' column shows how the evaluation results for each machine type and ID/Sec are aggregated to the machine-type level or the dev- and eval-subset level.
   $\mathrm{Amean}()$ and $\mathrm{Hmean}()$ are arithmetic mean and harmonic mean operations, respectively.
   $^*$: the attribute information is not available for some machine types.
   }
   \begin{adjustbox}{width=\textwidth}
   \begin{tabular}{rrrlllll}
      \toprule
      Year & \#MT & \#ID/Sec & Domain & Meta Information & Dev/Eval & Evaluation score & Aggregation \\
      \midrule
      2020~\cite{Koizumi_DCASE2020_01} & 6 & 6 or 7 & Source & MT, ID & ID & $\mathrm{Amean}(\mathrm{s\_auc}, \mathrm{s\_pauc})$ & $\mathrm{Amean}()$ \\
      2021~\cite{kawaguchi2021description} & 7 & 6 & Source, Target & MT, Sec, Attribute, Domain & Sec & $\mathrm{Hmean}(\mathrm{s\_auc}, \mathrm{s\_pauc}, \mathrm{t\_auc}, \mathrm{t\_pauc})$ & $\mathrm{Hmean}()$ \\
      2022~\cite{dohi2022description} & 7 & 6 & Source, Target & MT, Sec, Attribute, Domain & Sec & $\mathrm{Hmean}(\mathrm{smix\_auc}, \mathrm{tmix\_auc}, \mathrm{mix\_pauc})$ & $\mathrm{Hmean}()$ \\
      2023~\cite{dohi2023description} & 14 & 1 & Source, Target & MT, Sec, Attribute, Domain & MT & $\mathrm{Hmean}(\mathrm{smix\_auc}, \mathrm{tmix\_auc}, \mathrm{mix\_pauc})$ & $\mathrm{Hmean}()$ \\
      2024~\cite{nishida2024description} & 16 & 1 & Source, Target & MT, Sec, Attribute$^*$, Domain & MT & $\mathrm{Hmean}(\mathrm{smix\_auc}, \mathrm{tmix\_auc}, \mathrm{mix\_pauc})$ & $\mathrm{Hmean}()$ \\
      \bottomrule
      \end{tabular}
   \end{adjustbox}
   \vspace{-8pt}
   \label{tab:conditions}
\end{table*}

\subsection{Supported experimental conditions}
\label{sec:conditions}

ASDKit supports the DCASE 2020--2024 conditions~\cite{Koizumi_DCASE2020_01,kawaguchi2021description,dohi2022description,dohi2023description,nishida2024description}, which are summarized in Table~\ref{tab:conditions}.
Across these conditions, the machine types, the provided meta-information, and the evaluation scores differ.
ASDKit provides download scripts and evaluation tools for these datasets.
The download scripts automatically download the datasets and add ground-truth normal and anomalous labels to the test sets, for which the labels are concealed during the challenge and released by the organizers afterward.
The evaluation tool computes the official evaluation scores according to the respective DCASE year.
For all conditions, the datasets are divided into official \textit{dev} and \textit{eval} subsets, and evaluation scores are aggregated separately for each subset.
The evaluation score is based on a combination of several types of \ac{auc}.

In DCASE 2020, the machine ID, which identifies each individual machine of the same machine type, is provided as meta information.
The evaluation score is calculated as the arithmetic mean of the \ac{auc} and the \ac{pauc} with $p = 0.1$.
This score is computed for each machine ID within each machine type and then aggregated into dev- and eval-subset-level scores using the arithmetic mean.
The dev and eval subsets are defined based on the machine ID~\cite{Koizumi_DCASE2020_01}.

DCASE 2021 newly introduces a domain shift problem, where recording environments or machine operations differ between the \textit{source} and \textit{target} domains~\cite{wilkinghoff2025handling}.
While abundant training data is available in the source domain, only a few samples are available in the target domain.
As meta information, \textit{attributes} which denotes the recording and machine operation conditions, and the domain information are provided.
The evaluation score is computed as the harmonic mean of $\mathrm{s\_auc}$, $\mathrm{s\_pauc}$, $\mathrm{t\_auc}$, and $\mathrm{t\_pauc}$.
$\mathrm{s\_auc}$ and $\mathrm{s\_pauc}$ are the \ac{auc} and \ac{pauc} for the source domain, respectively, and $\mathrm{t\_auc}$ and $\mathrm{t\_pauc}$ are the \ac{auc} and \ac{pauc} for the target domain, respectively.
Therefore, \ac{asd} systems are expected to perform well in both domains.
Also, in DCASE 2021, the machine ID is replaced with the \textit{section}.
The section defines subsets within one machine type, and it serves a similar role to the machine ID; however, in some machine types, different machine IDs can appear in the same section, and the same machine ID can appear in multiple sections~\cite{kawaguchi2021description}.
The evaluation scores for each section within each machine type, are aggregated to dev- and eval-subset-level scores using the harmonic mean.

DCASE 2022 inherits the domain shift problem from the DCASE 2021 condition, but employs a different evaluation score~\cite{dohi2022description}.
The score is calculated as the harmonic mean of $\mathrm{smix\_auc}$, $\mathrm{tmix\_auc}$, and $\mathrm{mix\_pauc}$.
$\mathrm{smix\_auc}$ is the \ac{auc} calculated using normal and anomalous samples from the source domain together with anomalous samples from the target domain.
Similarly, $\mathrm{tmix\_auc}$ is the \ac{auc} calculated using normal and anomalous samples from the target domain together with anomalous samples from the source domain.
$\mathrm{mix\_pauc}$ is the \ac{pauc} calculated using normal and anomalous samples from both domains.
These evaluation scores are calculated by jointly using samples from both domains to assess whether the anomaly score distributions are well aligned between domains.

In DCASE 2023 and 2024, only one section is provided, but a larger number of machine types are included.
The dev and eval subsets are defined based on the machine types~\cite{dohi2023description,nishida2024description}.
In DCASE 2024, attribute information is not available for some machine types~\cite{nishida2024description}.
The evaluation score is the same as in DCASE 2022.

\begin{table}
   \centering
   \caption{ASD methods supported by ASDKit.}
   \begin{tabular}{ll}
      \toprule
      Approach & Recipe name \\
      \midrule
      AE & \textit{ae} \\
      Discriminative & \textit{dis\_spec\_adacos\_fixed}, \textit{dis\_beats\_scac\_trainable}, etc. \\
      Raw feature & \textit{raw\_spec}, \textit{raw\_beats}, \textit{raw\_eat} \\
      \bottomrule
      \end{tabular}
   \vspace{-8pt}
   \label{tab:supported}
\end{table}

\subsection{Supported methods}
ASDKit handles various \ac{asd} methods in a unified framework shown in Fig.~\ref{fig:framework}.
The supported methods are summarized in Table~\ref{tab:supported}.
All \ac{asd} methods consist of \textit{frontend} and \textit{backend} modules, where the frontend extracts some features from audio signals, and the backend computes anomaly scores after processing these features.
Training and evaluation are performed in four steps:
(1) training the frontend with the training dataset,
(2) extracting features from both the training and test datasets,
(3) training the backend with the features extracted from the training dataset and computing anomaly scores for the test data,
and (4) computing evaluation scores based on the anomaly scores and ground-truth normal and anomalous labels.
In the following, we explain the \ac{asd} methods supported by ASDKit and their processing pipelines within this four-step framework.

\subsubsection{AE}
\label{sec:ae}
This is the \ac{ae}-based official DCASE baseline method~\cite{Koizumi_DCASE2020_01}.
The frontend consists of a ten-layer \ac{mlp}.
The input features are five adjacent frames of the log Mel power spectrogram, and the frontend is trained to minimize their reconstruction loss.
\Ac{ae} directly computes anomaly scores based on the reconstruction error of encountered test samples in the second step, and in the third step, the backend simply copies these anomaly scores.
The \ac{ae} is independently trained for each machine type, repeating steps 1--4 for each machine type.


\subsubsection{Discriminative methods}
\label{sec:dis}
ASDKit provides various options for discriminative methods.
For the frontend architecture, it supports multi-branch CNNs~\cite{Wilkinghoff2023design,fujimura2025improvements} and \ac{ssl} models~\cite{zheng2024improving,pmlr-v202-chen23ag,ijcai2024p421}.
The multi-branch CNN receives multiple input features, such as the spectrum and spectrograms.
The frontend independently processes these features and obtains a final output by concatenating the outputs from each branch.
Each branch consists of 1D or 2D convolutional layers followed by an \ac{mlp}.
For \ac{ssl} models, ASDKit supports BEATs~\cite{pmlr-v202-chen23ag} and EAT~\cite{ijcai2024p421}, which are widely used in \ac{asd} tasks~\cite{zheng2024improving,jiang24c_interspeech,saengthong2025deep,jiang2025adaptive,yin2025diffusion}.
Based on previous works~\cite{zheng2024improving}, we also introduce additional \ac{lora}~\cite{hu2022lora} parameters to fine-tune these models.
These frontends are trained using classification of meta-information labels.

For the discriminative loss function, ASDKit supports ArcFace~\cite{deng2019arcface}, AdaCos~\cite{zhang2019adacos}, Sub-cluster AdaCos~(SCAC)~\cite{wilkinghoff2021sub}, and AdaProj~\cite{Wilkinghoff2024}, where these angular margin loss functions are known to be effective for \ac{asd} tasks~\cite{wilkinghoff2023angular}.
ASDKit also supports the option to choose between fixed and trainable class centers for the angular margin loss functions, as this choice affects performance; fixed class centers have been shown to achieve better results in previous work~\cite{Wilkinghoff2023design}.

For data augmentation, ASDKit supports Mixup~\cite{zhang2017mixup} and SpecAugment~\cite{park2019specaugment}, which are widely used in \ac{asd} tasks~\cite{Wilkinghoff2023design}.
Additionally, ASDKit supports techniques to boost \ac{asd} performance, such as FeatEx~\cite{wilkinghoff2024self} and its parameter-efficient variant, subspace loss~\cite{fujimura2025improvements}.
FeatEx and subspace loss introduce additional losses using subspace features in the multi-branch CNN, where subspace features refer to the features extracted from each branch before concatenation.

The backend consists of \ac{knn}, where anomaly scores are calculated as the average distance to the $k$ nearest neighbors in the training (reference) dataset from a given observation in the discriminative feature space.
ASDKit also supports variants that incorporate kmeans clustering~\cite{Wilkinghoff2023design}, SMOTE oversampling~\cite{chawla2002smote,jiang2025adaptive}, and anomaly score rescaling~\cite{wilkinghoff2025keeping}.
Kmeans clustering and SMOTE are used to balance the number of training samples between the source and target domains.
Kmeans clustering reduces the number of samples in the source domain, and the resulting centroids are used as reference samples in the subsequent \ac{knn}-based anomaly score calculation.
SMOTE is applied to oversample the training samples in the target domain.
The anomaly score rescaling technique~\cite{wilkinghoff2025keeping} rescales anomaly scores based on the local density of training samples in the feature space, addressing the tendency for low-density target domains to exhibit higher anomaly scores.

The discriminative methods train a common frontend using data from all machine types, and then independently train the backend for each machine type by repeating steps 2--4 for each type.

\begin{table}
   \centering
   \caption{Setups of the discriminative methods.
   MB indicates multi branch CNN.
   Parentheses in the ``Training strategy'' column denote whether the class centers in the loss function are fixed or trainable.}
   \begin{adjustbox}{width=0.95\columnwidth}
   \begin{tabular}{llll}
      \toprule
      Recipe Name & Frontend & \makecell[l]{Training\\strategy} & DataAug \\
      \midrule
      \makecell[l]{\textit{dis\_spec\_}\\\textit{adacos\_fixed\_wo\_mixup}} & MB & AdaCos (fixed) & No \\
      \midrule
      \textit{dis\_spec\_adacos\_fixed} & MB & AdaCos (fixed) & Mixup \\
      \midrule
      \textit{dis\_spec\_scac\_fixed} & MB & SCAC (fixed) & Mixup \\
      \midrule
      \textit{dis\_spec\_scac\_trainable} & MB & SCAC (trainable) & Mixup \\
      \midrule
      \textit{dis\_spec\_subspaceloss} & MB & Subspace loss & Mixup \\
      \midrule
      \textit{dis\_spec\_featex} & MB & FeatEx & Mixup \\
      \midrule
      \textit{dis\_multispec\_scac\_trainable} & MB & SCAC (trainable) & Mixup \\
      \midrule
      \textit{dis\_beats\_scac\_trainable} & \makecell[l]{BEATs\\w/ \acs{lora}} & SCAC (trainable) & Mixup \\
      \midrule
      \textit{dis\_eat\_scac\_trainable} & \makecell[l]{EAT\\w/ \acs{lora}} & SCAC (trainable) & Mixup \\
      \bottomrule
      \end{tabular}
   \end{adjustbox}
   \vspace{-4pt}
   \label{tab:discriminative_methods}
\end{table}

\subsubsection{Raw feature-based methods}
\label{sec:raw}
ASDKit supports raw feature-based \ac{asd} methods~\cite{guan2023time,saengthong2025deep}, where the frontend extracts raw features without training with meta-information label classification.
For example, as raw features, \cite{guan2023time} uses the time-averaged spectrogram, and \cite{saengthong2025deep} uses BEATs features without fine-tuning.
For the backend, the same \ac{knn}-based anomaly score calculation as in the discriminative methods is employed.
This method repeats steps 2--4 for each machine type, skipping the frontend training in the first step.


\begin{figure*}[t]
   \centering
   \centerline{\includegraphics[width=\textwidth]{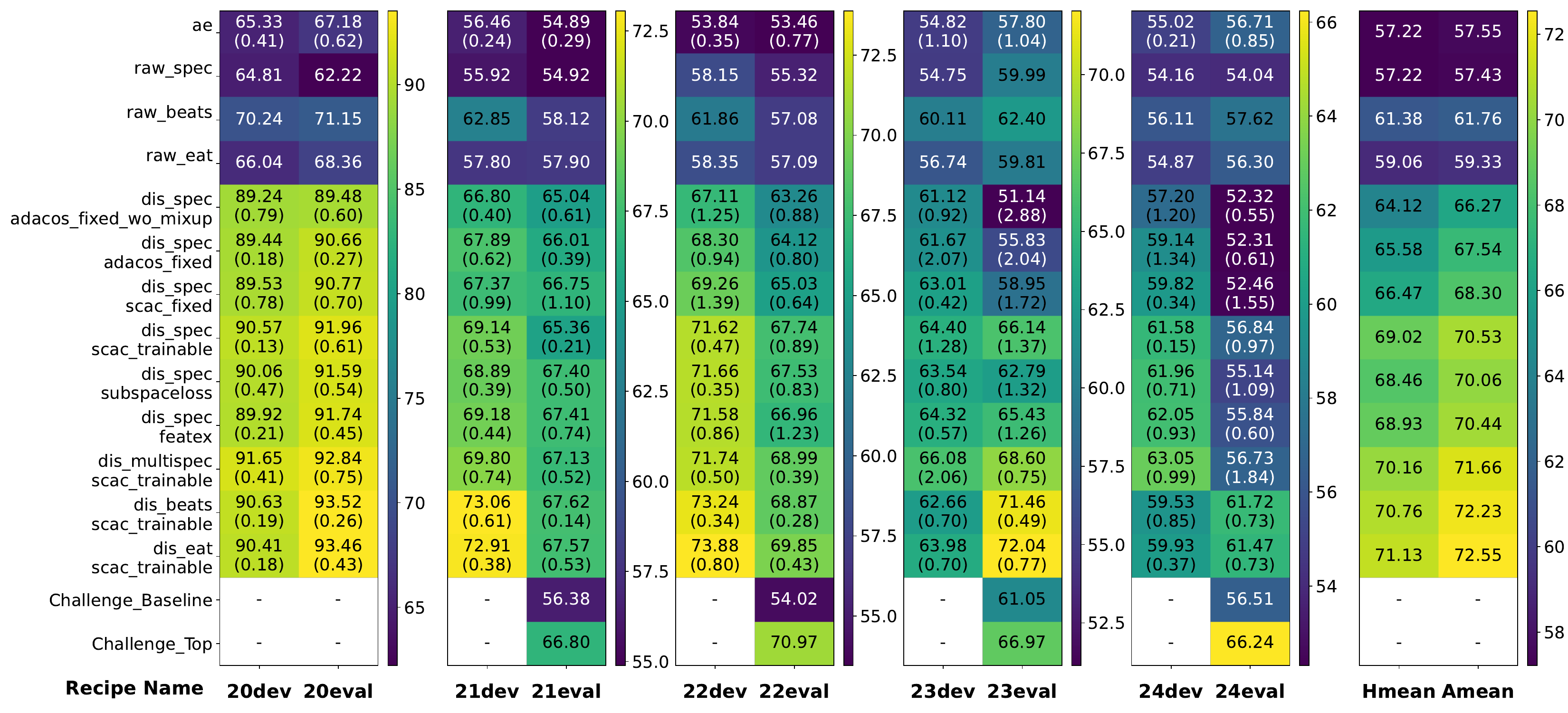}}
   \vspace{-8pt}
   \caption{Official evaluation scores. Values are presented as ``arithmetic mean (standard deviation)'' across four independent trials.
   Hmean and Amean denote the harmonic mean and arithmetic mean, respectively, computed over all dev and eval subsets across the datasets.
   Backend for the discriminative methods and raw feature-based methods is \ac{knn} with SMOTE.
   Raw feature-based methods do not involve any random processes; therefore, the standard deviation is not reported.
   }
   \vspace{-4pt}
   \label{fig:frontend}
   \end{figure*}

\section{Experimental Evaluation}
We re-evaluated various \ac{asd} methods using ASDKit to demonstrate its capabilities.

\subsection{Setups}
\subsubsection{Datasets and metrics}
We used the DCASE 2020--2024 datasets~\cite{Koizumi_DCASE2020_01,kawaguchi2021description,dohi2022description,dohi2023description,nishida2024description} and the official evaluation scores described in Sec.~\ref{sec:conditions}.
We computed the arithmetic mean and standard deviation across four independent trials, where each trial’s official score was calculated as the harmonic or arithmetic mean, as defined in Table~\ref{tab:conditions}.

\subsubsection{Evaluated methods and their setups}
In the following, we list the \ac{asd} methods evaluated in this paper and describe their setups.
Each method is referred to by its recipe name provided in ASDKit.

\textit{ae} is the \ac{ae}-based method~\cite{Koizumi_DCASE2020_01} described in Sec.~\ref{sec:ae}.
We trained the \ac{ae} network for 50 epochs using the Adam optimizer~\cite{Kingma_2015} using the mean squared error as a loss function with a fixed learning rate of 0.001 and a batch size of 256.
For the input features, we used five adjacent frames of the log Mel power spectrogram, with 128 Mel bins, a DFT size of 1024 samples, and a hop size of 512 samples.

The discriminative methods we evaluated are summarized in Table~\ref{tab:discriminative_methods}.
Mixup~\cite{zhang2017mixup} was always applied with a probability of 50\%.
\textit{dis\_multispec\_scac\_trainable} extracted 512-dimensional features from an amplitude spectrum and three spectrograms with different DFT sizes (256, 512, and 1024 samples), while the other multi-branch CNN methods extracted 256-dimensional features from an amplitude spectrum and an amplitude spectrogram with a DFT size of 1024 samples.
The hop size was set to half the DFT size, and frequency bins in the range of \SIrange{200}{8000}{\Hz} were used.
We trained the multi-branch CNNs for 16 epochs using AdamW optimizer~\cite{adamw2019} with a fixed learning rate of 0.001 and a batch size of 64.
For \textit{dis\_beats\_scac\_trainable}, we used the \textit{BEATs\_iter3.pt} checkpoint and introduced \ac{lora} parameters to the query and key projection layers within the Transformer.
The 768-dimensional feature sequence from BEATs was aggregated using a statistics pooling layer~\cite{desplanques2020ecapa} and then projected to a 256-dimensional feature using a linear layer.
For \textit{dis\_eat\_scac\_trainable}, we used the \textit{EAT-base\_epoch10\_pt.pt} checkpoint and introduced \ac{lora} parameters to the query, key, and value projection layers.
A 768-dimensional CLS feature from EAT was projected to a 256-dimensional feature using a linear layer.
We fine-tuned the \ac{ssl}-based models for 25 epochs using AdamW optimizer with a batch size of 8 and a \ac{lora} rank of 64.
The learning rate was linearly increased from 0 to 0.0001 over the first 5,000 steps.

For the raw feature-based methods described in Sec.~\ref{sec:raw}, we evaluated \textit{raw\_spec}, \textit{raw\_beats}, and \textit{raw\_eat}.
\textit{raw\_spec} used time-averaged Mel amplitude spectrograms with 128 Mel bins, a DFT size of 1024 samples, and a hop size of 512 samples.
\textit{raw\_beats} averaged the 768-dimensional feature sequence from BEATs, while \textit{raw\_eat} directly used the 768-dimensional CLS feature from EAT.

For the discriminative methods and raw feature-based methods, we used a \ac{knn}-based backend with $k = 1$ described in Sec.~\ref{sec:dis}.
We also used kmeans clustering, SMOTE oversampling, and anomaly score rescaling techniques.
For kmeans clustering, we set the number of clusters to 16.
For SMOTE oversampling, we set the oversampling ratio to 20\% and the number of neighbors to 2.
For anomaly score rescaling~\cite{wilkinghoff2025keeping}, we used a \ac{knn}-based approach with the number of neighbors set to 4.

\begin{figure*}[t]
   \centering
   \centerline{\includegraphics[width=0.98\textwidth]{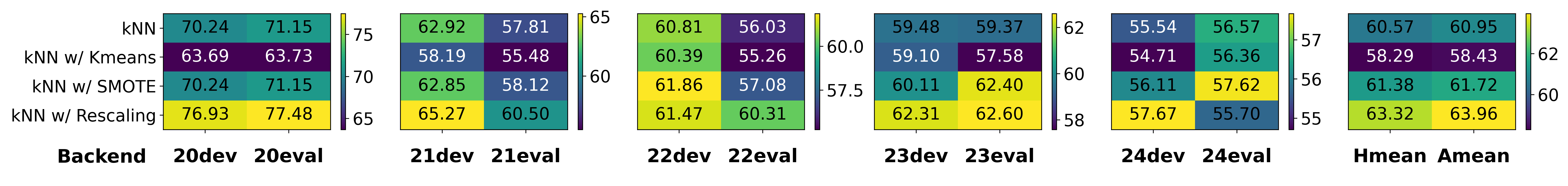}}
   \vspace{-14pt}
   \caption{Official evaluation scores of \textit{raw\_beats} with different backends.}
   \vspace{-4pt}
   \label{fig:backend_raw_beats}
   \end{figure*}
   
   \begin{figure*}[t]
   \centering
   \centerline{\includegraphics[width=0.98\textwidth]{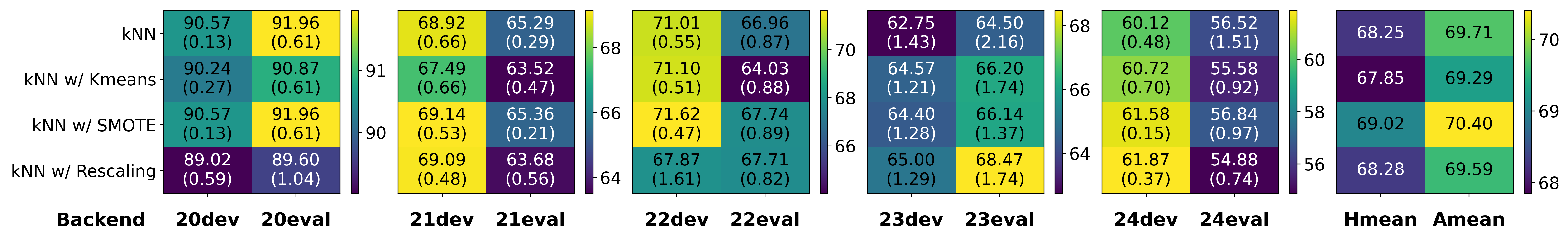}}
   \vspace{-10pt}
   \caption{Official evaluation scores of \textit{dis\_spec\_scac\_trainable} with different backends.
   Values are presented as ``arithmetic mean (standard deviation)'' across four independent trials.}
   \vspace{-8pt}
   \label{fig:backend_dis_spec}
   \end{figure*}
   

\subsection{Results}
Figure~\ref{fig:frontend} shows the official evaluation scores of the considered \ac{asd} methods, where both the discriminative methods and raw feature-based methods employ \ac{knn} with SMOTE as the backend.
The figure also includes the evaluation scores of the top-performing system and the official baselines reported by the DCASE organizers.\footnote{The evaluation scores on the development set are not officially reported. In DCASE 2020, the official scores aggregated across all machine types were also not provided.
Although several official baseline systems are provided but only the best-performing baseline systems are included in the figure. The best official baseline systems in DCASE 2021~\cite{Koizumi_DCASE2020_01} and 2024~\cite{nishida2024description} use the same method as our \textit{ae} recipe, while the 2023~\cite{dohi2023description} baseline is its variant~\cite{Harada_EUSIPCO2023_01}. The 2022 baseline employs a different discriminative method~\cite{dohi2022description}.}
First, it is evident that evaluation on a single dataset and a single trial is insufficient.
For example, the performance order of \textit{dis\_spec\_adacos\_fixed} and \textit{dis\_beats} is reversed between the DCASE 2023 dev and eval subsets.
Also, \textit{dis\_spec\_adacos\_fixed} exhibits a standard deviation of 2.04 on the DCASE 2023 eval subset, with performance ranging from 53.07 to 57.98 across four trials, as observed in our experimental results.

We reaffirm that training techniques for discriminative methods significantly impact performance.
For instance, the effectiveness of mixup is demonstrated by comparing \textit{dis\_spec\_adacos\_fixed\_wo\_mixup} and \textit{dis\_spec\_adacos\_fixed}; the effectiveness of SCAC, by comparing \textit{dis\_spec\_adacos\_fixed} and \textit{dis\_spec\_scac\_fixed}; and the effectiveness of multi-resolution spectrograms, by comparing \textit{dis\_spec\_scac\_trainable} and \textit{dis\_multispec\_scac\_trainable}.
Additionally, fine-tuned \ac{ssl} models (\textit{dis\_beats\_scac\_trainable} and \textit{dis\_eat\_scac\_trainable}) consistently achieve high performance.

As a new insight, we find that trainable class centers improve performance in most cases, as demonstrated by the comparison between \textit{dis\_spec\_scac\_fixed} and \textit{dis\_spec\_scac\_trainable}, while fixed class centers achieved better results in previous work~\cite{Wilkinghoff2023design}.
Moreover, the FeatEx (\textit{dis\_spec\_featex}) and subspace loss (\textit{dis\_spec\_subspaceloss}) techniques achieve performance improvements similar to those obtained by using trainable class centers.
Since both FeatEx and subspace loss employ trainable centers for additional loss terms, we conclude that their performance gains are mainly attributable to the use of trainable class centers.

Furthermore, Fig.\ref{fig:frontend} demonstrates that ASDKit achieves \ac{sota}-level performance.
Note that the top-performing systems in DCASE 2021~\cite{LopezIL2021}, 2022~\cite{LiuCQUPT2022}, and 2024~\cite{LvAITHU2024} employed ensembles, with the 2022 system additionally using machine-specific hyperparameter tuning, except for the 2023~\cite{JieIESEFPT2023} top system.

Figures~\ref{fig:backend_raw_beats} and \ref{fig:backend_dis_spec} show the official evaluation scores of \textit{raw\_beats} and \textit{dis\_spec\_scac\_trainable} with different backends, respectively.
The results show that \ac{knn} with SMOTE achieves high performance in most cases.
The anomaly score rescaling technique significantly improves the performance of \textit{raw\_beats}, although our re-evaluation across multiple datasets reveals that its effect is unstable for the considered discriminative embedding model.

\section{Conclusion}
In this paper, we introduced ASDKit, an open-source toolkit for \ac{asd} research.
ASDKit provides recipes for various \ac{asd} methods, including \ac{ae}-based approaches, state-of-the-art (SOTA) discriminative methods, and raw feature-based methods.
In addition, it supports comprehensive evaluation on the DCASE 2020--2024 datasets with multiple trials.
Using ASDKit, we re-evaluated various \ac{asd} methods and identified the effectiveness of mixup, SCAC with trainable class centers, multi-resolution spectrograms, and fine-tuned \ac{ssl} models.
We will continuously update ASDKit with new \ac{asd} methods and welcome contributions from the community to facilitate further development of \ac{asd} research.

\section{Acknowledgment}
\label{sec:ack}
This work was partly supported by JSPS KAKENHI Grant Number JP25KJ1439.


\clearpage
\section*{References}
\printbibliography







\end{document}

%% file: settings.tex
\usepackage{amssymb}
\usepackage{multirow}
\usepackage{makecell}
\usepackage{bm}
\usepackage{booktabs}
\usepackage{comment}
\usepackage{pifont}
\usepackage{color}
\usepackage{siunitx}
\usepackage{amsfonts}
\usepackage{adjustbox}
\usepackage{listings}
\usepackage{xcolor}

\definecolor{codegray}{rgb}{0.5,0.5,0.5}
\definecolor{codepurple}{rgb}{0.58,0,0.82}
\definecolor{backcolour}{rgb}{0.95,0.95,0.92}

\usepackage{listings}
\usepackage{xcolor}

\lstdefinestyle{mystyle}{
    backgroundcolor=\color{white}, 
    basicstyle=\ttfamily\footnotesize,
    breaklines=true,
    numbers=none, 
    showstringspaces=false,
    showtabs=false,
    frame=single,
    keywordstyle=\color{blue},
    commentstyle=\color{gray},
    stringstyle=\color{orange}
}

\usepackage[backend=biber,style=ieee,
maxbibnames=6,
doi=false,isbn=false,url=false,eprint=false
]{biblatex}
\defbibheading{bibliography}[\refname]{}

\DeclareSourcemap{
	\maps[datatype=bibtex, overwrite=true]{
		\map{
		    \step[fieldsource=booktitle,
			match=\regexp{.*EUSIPCO.*},
			replace={Proc. EUSIPCO}]
			\step[fieldsource=booktitle,
			match=\regexp{.*CVPR.*},
			replace={Proc. CVPR}]
			\step[fieldsource=booktitle,
			match=\regexp{.*Interspeech.*},
			replace={Proc. Interspeech}]
			\step[fieldsource=booktitle,
			match=\regexp{.*ICASSP.*},
			replace={Proc. ICASSP}]
			\step[fieldsource=booktitle,
			match=\regexp{.*ICLR.*},
			replace={Proc. ICLR}]
			\step[fieldsource=booktitle,
			match=\regexp{.*IJCAI.*},
			replace={Proc. IJCAI}]
			\step[fieldsource=booktitle,
			match=\regexp{.*IJCNN.*},
			replace={Proc. IJCNN}]
			\step[fieldsource=booktitle,
			match=\regexp{.*ICML.*},
			replace={Proc. ICML}]
			\step[fieldsource=booktitle,
			match=\regexp{.*ASRU.*},
			replace={Proc. ASRU}]
			\step[fieldsource=booktitle,
			match=\regexp{.*SLT.*},
			replace={Proc. SLT}]
			\step[fieldsource=booktitle,
			match=\regexp{.*SSW.*},
			replace={Proc. SSW}]
			\step[fieldsource=booktitle,
			match=\regexp{.*WASPAA.*},
			replace={Proc. WASPAA}]
			\step[fieldsource=booktitle,
			match=\regexp{.*DCASE.*},
			replace={Proc. DCASE}]
            \step[fieldsource=booktitle,
            match=\regexp{.*Detection.*and.*Classification.*of.*Acoustic.*Scenes.*and.*Events.*Workshop.*},
			replace={Proc. DCASE}]
            \step[fieldsource=booktitle,
			match=\regexp{.*MLSP.*},
			replace={Proc. MLSP}]
            \step[fieldsource=booktitle,
			match=\regexp{.*ECCV.*},
			replace={Proc. ECCV}]
            \step[fieldsource=journal,
			match=\regexp{.*NeurIPS.*},
			replace={Advances in NeurIPS}]
            \step[fieldsource=journal,
			match=\regexp{.*TASLP.*},
			replace={IEEE/ACM TASLP}]
            \step[fieldsource=journal,
			match=\regexp{.*JAIR.*},
			replace={JAIR}]
            \step[fieldsource=journal,
			match=\regexp{.*J-STSP.*},
			replace={IEEE J-STSP}]                
			\step[fieldsource=series,
			match=\regexp{.+},
			replace={{}}]
			\step[fieldsource=editor,
			match=\regexp{.+},
			replace={{}}]
			\step[fieldsource=publisher,
			match=\regexp{.+},
			replace={{}}]
			\step[fieldsource=month,
			match=\regexp{.+},
			replace={{}}]
			\step[fieldsource=location,
			match=\regexp{.+},
			replace={{}}]
			\step[fieldsource=address,
			match=\regexp{.+},
			replace={{}}]
			\step[fieldsource=organization,
			match=\regexp{.+},
			replace={{}}]
		}
	}
}